\documentclass[sigconf,nonacm,natbib=true]{acmart}

\AtBeginDocument{%
}

\usepackage[linesnumbered,ruled,vlined,noend]{algorithm2e}

\DontPrintSemicolon
\SetAlgoNlRelativeSize{-2}
\SetKwInput{Require}{Input}
\SetKwInput{Ensure}{Output}
\SetKw{Return}{Return}

\usepackage{enumitem}
\usepackage{multirow}
\usepackage{pbalance}
\usepackage{makecell}
\usepackage{soul}       
\usepackage{pifont}
\usepackage{fontawesome5}

\setcitestyle{numbers,sort&compress}

\newcommand{\uline}[1]{\ul{#1}}

\newcommand{\floatnote}[1]{%
  \vspace{2pt}%
  {\footnotesize\itshape
   \setlength{\parskip}{0pt}%
   \setlength{\parindent}{0pt}%
   #1\par}%
}

\newcommand{\thdr}[2]{\shortstack[c]{#1\\[-2pt]\footnotesize #2}}

\newcommand{\cmark}{\ding{51}}

\newlist{rqlist}{itemize}{1}
\setlist[rqlist]{%
  label={},
  leftmargin=*,
  itemsep=1pt,
  topsep=2pt,
  parsep=0pt,
  partopsep=0pt
}

\newcommand{\systemname}{\textsc{NuggetIndex}}

\newsavebox{\promptbox}

\newenvironment{prompt}{%
  \par\addvspace{6pt}%
  \noindent
  \begingroup
  \setlength{\fboxrule}{0.4pt}%
  \setlength{\fboxsep}{0pt}%
  \begin{lrbox}{\promptbox}%
  \begin{minipage}{\dimexpr\linewidth-2\fboxrule\relax}%
    \vspace*{6pt}%
    \noindent\hspace*{8pt}%
    \begin{minipage}{\dimexpr\linewidth-16pt\relax}%
      \setlength{\parindent}{0pt}%
}{%
    \end{minipage}%
    \par\vspace*{6pt}%
  \end{minipage}%
  \end{lrbox}%
  \fcolorbox{gray!50}{gray!8}{\usebox{\promptbox}}%
  \endgroup
  \par\addvspace{6pt}%
}

\begin{document}

\title{\systemname{}: Governed Atomic Retrieval for Maintainable RAG}

\author{Saber Zerhoudi}
\orcid{0000-0003-2259-0462}
\affiliation{%
  \institution{University of Passau}
  \city{Passau}
  \country{Germany}
}
\email{szerhoudi@acm.org}

\author{Michael Granitzer}
\orcid{0000-0003-3566-5507}
\affiliation{%
  \institution{Interdisciplinary Transformation University Austria}
  \city{Linz}
  \country{Austria}
}
\affiliation{%
  \institution{University of Passau}
  \city{Passau}
  \country{Germany}
}
\email{michael.granitzer@uni-passau.de}

\author{Jelena Mitrovi\'{c}}
\orcid{0000-0003-3220-8749}
\affiliation{%
  \institution{University of Passau}
  \city{Passau}
  \country{Germany}
}
\email{jelena.mitrovic@uni-passau.de}

% --- ACM-like first-page rights styling ---
\makeatletter
% Slightly tighten footnote area spacing (like acmart)
\setlength{\skip\footins}{9pt plus 2pt minus 1pt}

% Font & layout for the rights block (≈ 8pt on 9.5pt leading)
\newcommand{\acmrightssize}{\fontsize{8}{9.5}\selectfont}

\setlength{\emergencystretch}{1.5em} % try 1em–3em

% keep this
\settopmatter{printacmref=false}

% Unnumbered, ACM-styled first-page footnote helper
\newcommand{\firstpagerights}[1]{%
  \begingroup
    \renewcommand\thefootnote{}%
    \footnotetext{%
      \acmrightssize
      \raggedright
      \setlength{\parskip}{0pt}%
      \setlength{\parindent}{0pt}%
      #1%
    }%
    \addtocounter{footnote}{0}%
  \endgroup
}
\makeatother

\begin{abstract}
Retrieval-augmented generation (RAG) systems are frequently evaluated via fact-based metrics, yet standard implementations retrieve passages or static propositions. This unit mismatch between evaluation and retrieval objects hinders maintenance when corpora evolve and fails to capture superseded facts or source disagreements. We propose \textsc{NuggetIndex}, a retrieval system that stores atomic information units as managed records, so called nuggets. Each record maintains links to evidence, a temporal validity interval, and a lifecycle state. By filtering invalid or deprecated nuggets prior to ranking, the system prevents the inclusion of outdated information. We evaluate the approach using a nuggetized MS~MARCO subset, a temporal Wikipedia QA dataset, and a multi-hop QA task. Against passage and unmanaged proposition retrieval baselines, \systemname{} improves nugget recall by 42\%, increases temporal correctness by 9 percentage points without the recall collapse observed in time-filtered baselines, and reduces conflict rates by 55\%. The compact nugget format reduces generator input length by 64\% while enabling lightweight index structures suitable for browser-based and resource-constrained deployment. We release our implementation, datasets, and evaluation scripts\footnote{\href{https://github.com/searchsim-org/sigir26-nuggetindex}{https://github.com/searchsim-org/sigir26-nuggetindex}}.
\end{abstract}




\begin{teaserfigure}
  \begin{center}
\includegraphics[width=0.95\textwidth]{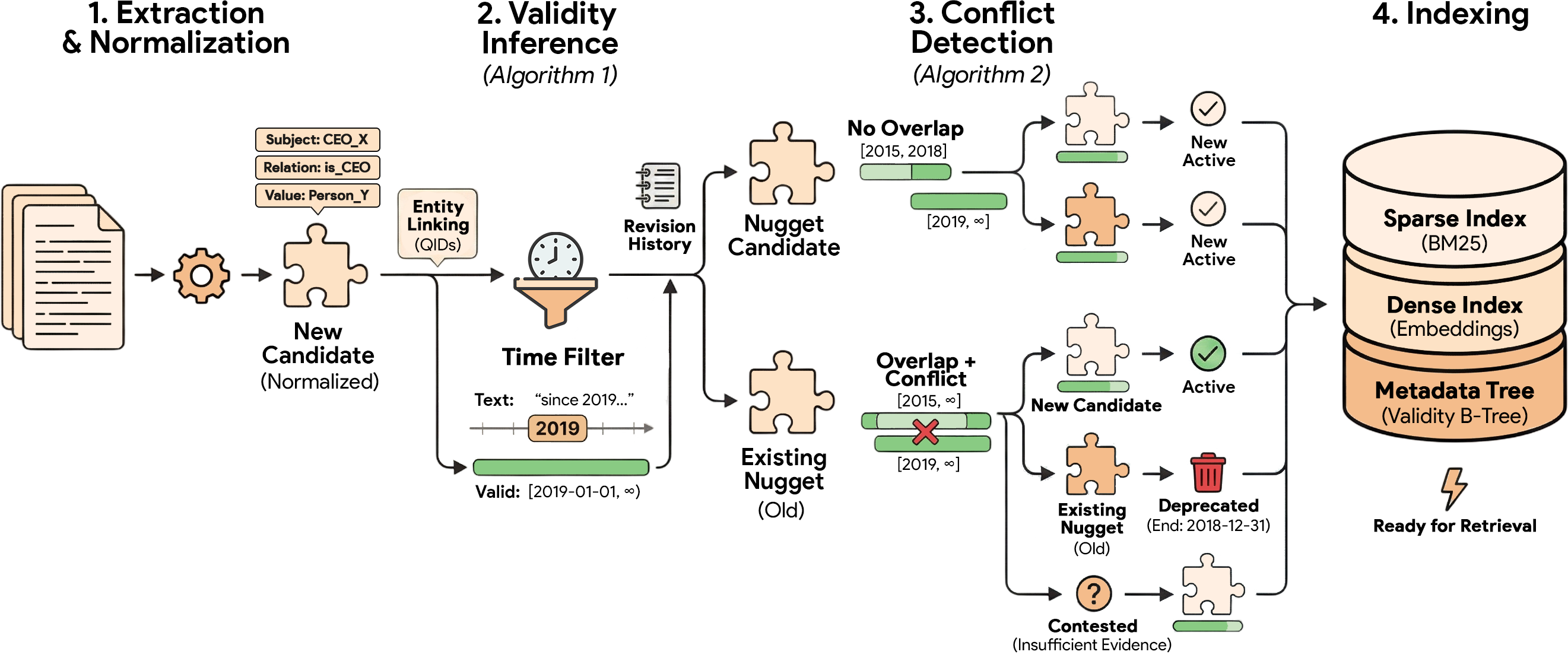}
% \vspace{-3mm}
  \caption{\systemname{} pipeline. Raw text is normalized into atomic candidates. Algorithm~\ref{alg:infer-validity} infers validity intervals using temporal expressions and revision history, while Algorithm~\ref{alg:conflict-detection} detects conflicts with the index to determine lifecycle states.}
\label{fig:pipeline}
  % \vspace{3mm}
\end{center}
\end{teaserfigure}

\maketitle
\enlargethispage{2\baselineskip}
\firstpagerights{%
  © ACM, 2026. This is the author's version of the work.\\
  The definitive version was published in:
  \emph{Proceedings of the 49th International ACM SIGIR Conference on Research and Development in Information Retrieval (SIGIR '26), July 20--24, 2026, Melbourne, VIC, Australia}.\\
  DOI: \url{https://doi.org/10.1145/3805712.3809687}
}

\section{Introduction}
\label{sec:intro}

Retrieval-augmented generation (RAG) grounds large language model (LLM) outputs in external evidence to reduce hallucination~\cite{lewis2020retrieval}. A primary challenge for deployed RAG systems is maintaining correctness as the underlying corpus evolves~\cite{huwiler2025versionrag,parry2025shelflife}.

In dynamic domains such as software documentation, news, or law, updates occur frequently~\cite{chen2021dataset}. Standard passage-based systems may retrieve conflicting information from different versions of a document. The generator might then merge outdated and current details, producing confident but incorrect answers. This issue arises from retrieving valid-looking but temporally inconsistent evidence.

Recent research suggests that retrieving finer-grained units, such as propositions or triplets, improves relevance and reduces token costs~\cite{chen2024dense}, and that augmenting RAG with user-centric agents yields further gains on personalised tasks~\cite{zerhoudi2024personarag}. However, existing implementations store these units as static text or embeddings, lacking metadata to capture temporal validity or source disagreement.

Simultaneously, evaluation methodologies are shifting toward nugget-based scoring, where a ``nugget'' is defined as an atomic, self-contained fact~\cite{voorhees2003overview, pradeep-etal-2024-autonuggetizer}. This trend creates a design misalignment: systems are evaluated on precise atomic facts but retrieve unmanaged, multi-fact passages. Consequently, retrieval systems lack mechanisms to explicitly track or update the specific facts required for accurate generation.

We propose bridging this gap with \systemname{}, an AI-governed index where retrieval units are not merely strings, but \textbf{versioned records with epistemic annotations}. We formally define a nugget not as a text span, but as a composite record $N = (f, v, e)$, containing the atomic fact $f$, its temporal validity $v$, and its epistemic status $e$ (e.g., agreed vs.\ contested). 

This data model relies on three properties to ensure maintenance. First, every nugget is assigned a \textit{temporal validity interval}~\cite{snodgrass2012tsql2}, $[t_s, t_e)$, defining when the fact is true, rather than when it was written. Second, we explicitly model epistemic uncertainty through a lifecycle state: \textsf{Active} for current consensus, \textsf{Deprecated} for superseded facts, and \textsf{Contested} for unresolved disagreement. Third, nuggets maintain provenance links to source spans, grounding these validity judgments in evidence.

At query time, \systemname{} filters the index to retain only nuggets that are \textsf{Active} and valid for the specific query time. \textsf{Deprecated} nuggets are excluded, while \textsf{Contested} items can be explicitly flagged. This filtering occurs before ranking, ensuring the generator processes only consistent and valid evidence.

We make the following contributions:
\begin{enumerate}[label=(\arabic*),leftmargin=2em,labelsep=0.4em]
    \item We formalize RAG maintenance as a retrieval task and identify the discrepancy between nugget evaluation and retrieval.
    \item We introduce the \systemname{} data model, which integrates validity intervals and lifecycle states, alongside algorithms for index construction and conflict resolution.
    \item We propose a governed retrieval pipeline that filters by temporal validity and lifecycle states before ranking. Ablations reveal that sparse retrieval outperforms dense on atomic nuggets, enabling sub-millisecond latency without embedding inference.
    \item We evaluate \systemname{} across four benchmarks, demonstrating a 42\% increase in nugget recall, 9.1 percentage point improvement in temporal correctness, and 55\% reduction in conflict rates, while reducing generator input by 64\%.
\end{enumerate}

\section{Background and Related Work}
\label{sec:related}

This section reviews four research areas relevant to our study: retrieval granularity, nugget-based evaluation, temporal information retrieval, and structured knowledge. We position \systemname{} relative to these approaches at the end of the section.

\subsection{Retrieval Granularity}

Standard RAG systems typically retrieve passages ranging from 100 to 200 tokens and concatenate them to form a prompt for the generator~\cite{lewis2020retrieval,karpukhin2020dense,izacard2021leveraging}. While this approach is straightforward, it suffers from inherent limitations. Passages frequently contain a mixture of relevant and irrelevant content, and multiple passages may redundantly state the same information. Furthermore, long contexts increase the computational cost of generation~\cite{liu2023lost}.

To address these issues, recent research has explored retrieving finer-grained units. For instance, \citet{chen2024dense} extract propositions—single-sentence atomic statements—and perform retrieval at this level, reporting improved precision. Similarly, triplet-based approaches extract subject-relation-object tuples and retrieve over these structures~\cite{gong2025beyond}. These methods succeed in improving relevance density and reducing prompt length. However, such atomic units are generally stored as plain text strings or embeddings without additional structure. They lack metadata indicating when a fact is valid or whether it has been superseded. Consequently, if two propositions contradict one another, the system lacks the mechanism to detect or resolve the conflict, potentially retrieving both and passing the contradiction to the generator.

\subsection{Nugget-Based Evaluation}

Nugget-based evaluation originated in the TREC QA tracks, where system answers were scored based on their coverage of ``information nuggets''—atomic facts that a complete answer should contain~\cite{voorhees-2003-trecqa, pavlu-etal-2012-nugget-eval}. This method offers greater robustness than string matching because it credits semantically equivalent answers even when their surface forms differ.

Recent advancements have scaled nugget-based evaluation through the use of LLM-based autograders~\cite{pradeep-etal-2024-autonuggetizer}. Given a candidate answer and a set of gold nuggets, an autograder determines which nuggets the answer successfully covers, enabling evaluation on large benchmarks without requiring manual annotation for every response. The reliance on nugget coverage offers an opportunity to align retrieval design with evaluation by targeting nuggets directly. This alignment aids in diagnosing failures and enables the optimization of retrieval models for nugget recall.

A parallel line of work uses nugget-like atomic units inside the retrieval pipeline rather than only at evaluation time. GINGER~\cite{lajewska2025ginger} grounds response generation in atomic information nuggets retrieved per query; proposition-style retrievers~\cite{chen2024dense} index single-sentence propositions and report precision gains on factoid QA; triplet-driven retrievers~\cite{gong2025beyond} index subject--relation--object tuples for RAG. These approaches share our atomic-unit choice but treat units as static text, omitting the validity interval and lifecycle state we maintain. On the temporal knowledge graph side, hybrid structural--semantic frameworks~\cite{deng2025multiexpert} model the evolution of relations between entities over time and answer queries against curated graphs that combine structural and semantic signals to capture historical patterns; these systems rely on a fixed schema and pre-existing graph structure. \systemname{} sits between these two lines: we adopt the atomic-unit retrieval substrate of nugget-IR while importing the validity-interval discipline of temporal KGs into the retrieval index itself, with the schema discovered from the target corpus rather than authored in advance.

\subsection{Temporal Information Retrieval}

Temporal Information Retrieval (IR) incorporates time into retrieval processes. Document-level methods use publication timestamps to rank by recency or filter within time windows~\cite{li-croft-2003-time}. Query-level approaches identify temporal intent to adjust retrieval strategies~\cite{berberich2010language}.

In the context of RAG, the core challenge is \emph{fact-level} temporal validity. A document published in 2023 may discuss events from 2020, while a document from 2018 may state a fact that remains true until 2022. Document timestamps do not capture these nuances; validity information is required at the fact level. Temporal question answering benchmarks evaluate whether systems provide answers appropriate for a specific query time~\cite{chen2021dataset}. Systems that retrieve mixed-time evidence typically perform poorly on these evaluations.

\subsection{Structured Knowledge for RAG}

Knowledge bases like Wikidata store facts with qualifiers such as temporal constraints and references~\cite{wikidata-datamodel,vrandevcic2014wikidata}. Wikidata statements can also be marked with ranks, such as preferred or deprecated, to indicate currency. Although this model is conceptually close to our proposed approach, Wikidata relies on a fixed ontology and formal schema. Populating the base requires entity linking and relation extraction against that specific schema, and querying requires structured languages like SPARQL~\cite{sparql11-query}. These requirements limit the applicability of such systems to open-domain text corpora where facts do not conform to a predefined structure.

Alternatively, GraphRAG approaches construct graphs over documents and utilize graph traversal or graph neural networks for evidence retrieval~\cite{knollmeyer2025document}. Although these methods enable multi-hop reasoning by capturing relationships, they require expensive graph construction and suffer from increased inference latency. Industrial systems such as AWS Neptune's Lexical Graph~\cite{aws2024lexicalgraph} hybridize lexical and graph indices to reduce that cost, but, like GraphRAG, they treat retrieved evidence as static and surface no temporal-validity, conflict, or rename signals to the generator.

\subsection{Positioning of \systemname{}}

\begin{table}[t]
\caption{Comparison of retrieval approaches for RAG.}
\label{tab:positioning}
\centering
\small
\renewcommand{\arraystretch}{1.15}
\setlength{\tabcolsep}{3.5pt}

\begin{tabular*}{\columnwidth}{@{}l@{\extracolsep{\fill}}ccccc@{}}
\toprule
& \multicolumn{1}{c}{\textbf{Gran.}} 
& \multicolumn{2}{c}{\textbf{Temporal}} 
& \multicolumn{1}{c}{\textbf{Baseline}} 
& \multicolumn{1}{c}{\textbf{Eff.}} \\
\cmidrule(lr){2-2}\cmidrule(lr){3-4}\cmidrule(lr){5-5}\cmidrule(lr){6-6}
\textbf{Method} &
\textbf{AU} &
\textbf{VI} &
\textbf{LS} &
\textbf{IR} &
\textbf{Cost} \\
\midrule
Passage RAG            & -- & -- & -- & \cmark & \cmark \\
Proposition RAG        & \cmark & -- & -- & \cmark & \cmark \\
Wikidata + retrieval   & \cmark & \cmark & \cmark & -- & -- \\
GraphRAG               & -- & -- & -- & -- & -- \\
\textsc{NuggetIndex}   & \cmark & \cmark & \cmark & \cmark & \cmark \\
\bottomrule
\end{tabular*}

% \vspace{-4mm}
\floatnote{{Gran.} granularity; \textit{Eff.} efficiency.
\textit{AU} atomic units; \textit{VI} validity intervals; \textit{LS} lifecycle states; \textit{IR} standard IR; \textit{Cost} low query cost. \cmark\ = supported; -- = not supported.}
% \vspace{-2mm}
\end{table}

Table~\ref{tab:positioning} compares \systemname{} to prior approaches, with the main distinction being the integration of validity data.

\textbf{vs.\ GraphRAG:} GraphRAG~\cite{knollmeyer2025document} emphasizes structural connectivity and multi-hop traversals. These implementations typically treat edges as static, lacking mechanisms to track temporal status or disputes. \systemname{} prioritizes \textit{validity governance} over topological complexity, ensuring retrieved facts are temporally accurate.

\textbf{vs.\ Wikidata:} Wikidata models validity using a ranking system (\textsc{\small Preferred}, \textsc{\small Normal}, \textsc{\small Deprecated})~\cite{vrandevcic2014wikidata}. This relies on manual curation and rigid ontologies. \systemname{} approximates this model via automated extraction from unstructured text, creating a scalable, \textit{probabilistic Wikidata} that operates without human maintenance.

In summary, \systemname{} defines the \textit{validity-scoped fact} as the fundamental unit of automated retrieval.

\section{\systemname{}: Governed Nugget Retrieval}
\label{sec:method}

\begin{figure}[H]
\centering
\includegraphics[width=\columnwidth]{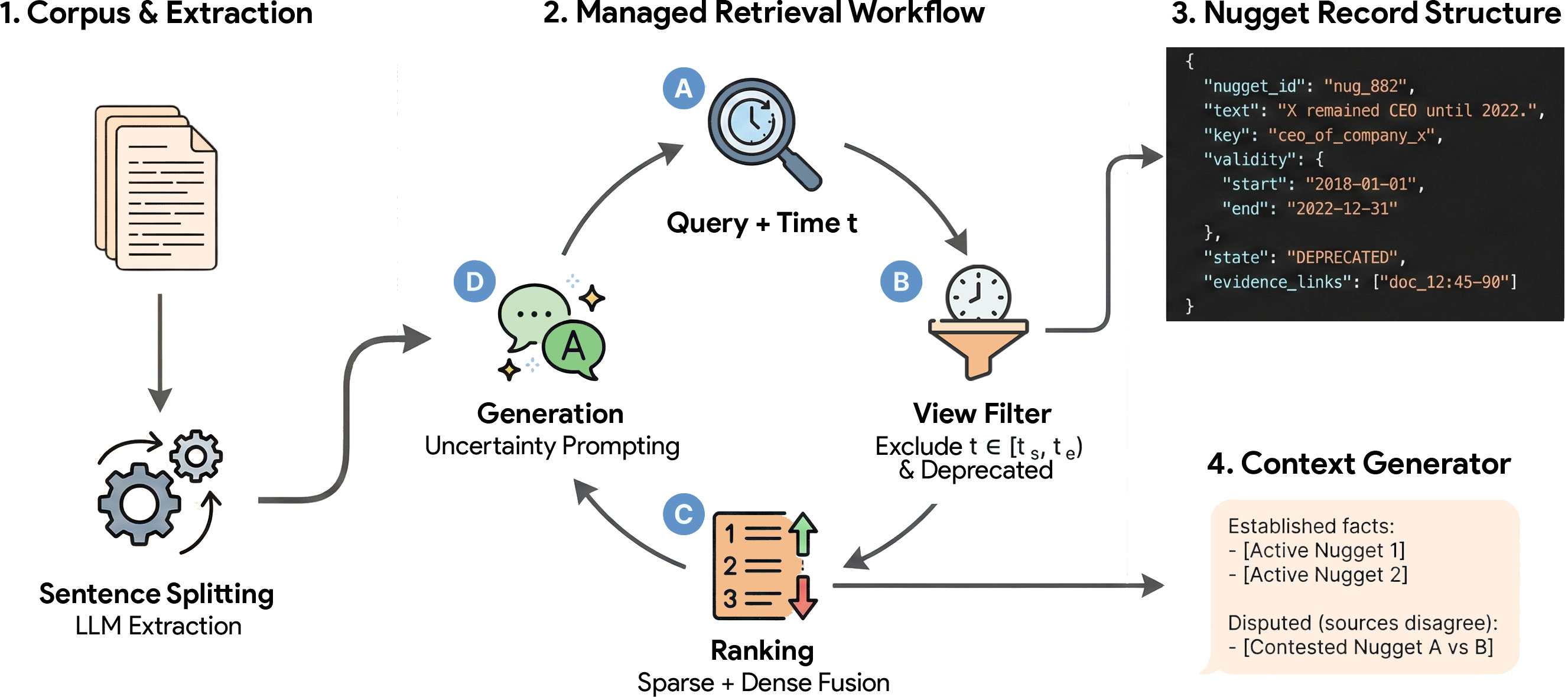}
\captionsetup{type=figure}
\vspace{-2em}
\caption{The \systemname{} architecture.}
\label{fig:architecture}
\floatnote{Documents are processed into atomic nuggets with temporal validity intervals and lifecycle states. At query time, the system filters by validity and state before ranking.}
\vspace{-2mm}
\end{figure}

\noindent\textbf{Operational scope and terminology.}
We use \emph{fact} to denote a (subject, predicate, object) triple extracted from a document span; we do not address philosophical truth or epistemological certainty. \emph{Validity} is the time interval during which a source asserts the fact, recovered from explicit temporal expressions when present (Algorithm~\ref{alg:infer-validity}); when absent, the interval is anchored to the document timestamp and flagged as inferred. \emph{Consensus} is operationalized as agreement of $\geq 2$ independent sources on the same canonicalized key; it is corpus-relative, not absolute. A fact is \textsc{Contested} when two or more sources commit incompatible objects on the same functional-predicate key over overlapping validity, with neither side reaching the evidence threshold. The system therefore targets RAG maintenance over corpora with extractable temporal signal and verifiable evidence; deep historical uncertainty (e.g.\ disputes over events with no contemporary record) and adversarial corpora (where coordinated misinformation can satisfy the evidence threshold) are out of scope and require human-in-the-loop verification, which the released annotation tool supports.

\subsection{Nugget Record Model}

A nugget $n$ is defined as a tuple $(k, f, \sigma, \epsilon, \Pi)$:
\begin{itemize}[label=\textendash,labelindent=0pt,leftmargin=2em,labelsep=0.4em]
    \item $k$: Nugget kind $\in$ $\{\textsc{\small SemanticFact},\ \textsc{\small EpisodicEvent},\ \textsc{\small Instruction}, \\ \textsc{\small UserPreference}\}$.
    \item $f$: Atomic fact as a semantic triple (\textit{subject}, \textit{predicate}, \textit{object}).
    \item $\sigma$: Temporal validity interval $[t_s, t_e)$ with scope, location, and source type.
    \item $\epsilon$: Epistemic state with status (\textsc{\small Active}, \textsc{\small Deprecated}, \textsc{\small Contested}), rank (\textsc{\small Preferred}, \textsc{\small Normal}, \textsc{\small Deprecated}), and confidence score.
    \item $\Pi$: Provenance record with evidence links to source spans, creation metadata, and version lineage.
\end{itemize}

Each nugget has a stable hash-based identifier and version graph links via a parent ID. The nugget key $\kappa = (s_{\text{norm}}, p, \textit{scope})$ enables conflict detection. A nugget is retrievable at query time $t$ if $t \in [t_s, t_e)$ and its status is \textsc{\small Active} (or \textsc{\small Contested} when uncertainty markers are desired). This differs from passage retrieval, which returns text blocks, and GraphRAG, which traverses entity graphs.

\subsection{Nugget Construction Pipeline}
\label{sec:construction}

Given a document $d$ with timestamp $t_d$, the extraction process involves four stages (Figure~\ref{fig:architecture}). Algorithm~\ref{alg:infer-validity} details validity inference, while Algorithm~\ref{alg:conflict-detection} outlines conflict detection.

\subsubsection*{Stage 1: Candidate Extraction}
We segment the document $d$ into sentences and process them using a sliding context window (the current sentence plus the preceding one) to preserve local context for pronoun resolution and cross-sentence references (e.g., ``He founded the company''). An LLM-based extractor decomposes each window into atomic statements. Each extracted candidate consists of the text $\ell$, extracted fields (\textit{subject}, \textit{predicate}, \textit{object}), and character offsets representing the evidence span in the source document.

\subsubsection*{Stage 2: Canonicalization and Keying}
We canonicalize extracted triples to support deduplication. Subjects are normalized through alias lookup (e.g., ``USA'' $\to$ ``United States'') and case standardization. Predicates map to a controlled vocabulary (e.g., ``is the CEO of'' $\to$ \texttt{\small chiefExecutiveOfficer}). Objects are resolved to entities where possible, or normalized to standard formats (e.g., ISO dates). The controlled vocabulary is a user-replaceable schema declaring, for each predicate, a canonical name, surface-form aliases, expected (subject, object) entity types, and a cardinality flag (functional, multi-valued, or event-log) that the conflict detector reads. To lower the adoption barrier on unfamiliar corpora, an opt-in schema-discovery module scans a representative sample, clusters surface-form predicates by frequency and entity-type co-occurrence, and proposes a starter schema that the user can edit before ingest.

We then compute the nugget key $\kappa = (s_{\text{norm}}, p, \textit{scope})$, where \textit{scope} is derived from document metadata or textual cues. The default is \texttt{global}, while \texttt{user} and \texttt{group} denote personal and shared contexts. Excluding the object value from the key enables the identification of duplicates (same key, similar values) and conflicts (same key, different values).

\subsubsection{Stage 3: Validity Inference}
We assign a validity interval $\sigma_n = [t_s, t_e)$ following Algorithm~\ref{alg:infer-validity}. A temporal tagger (e.g., SUTime~\cite{chang2012sutime} or regex patterns) identifies date expressions and classifies them as start points, end points, or point-in-time references. The start time $t_s$ uses an explicit start date if found (e.g., ``since 2019''), otherwise defaulting to the document timestamp $t_d$. The end time $t_e$ uses an explicit end date if found, otherwise defaulting to open-ended ($+\infty$).

For versioned sources such as Wikipedia, we refine intervals using revision history. If a statement in revision $r_i$ is absent in $r_{i+1}$, we set $t_e$ to the earlier of its current value and $r_{i+1}$'s timestamp. If a statement first appears in $r_{i+1}$, we set $t_s$ to the later of its current value and $r_{i+1}$'s timestamp. We assume revisions reflect real-world state changes. While suitable for collaborative sources, editorial corrections risk being interpreted as temporal transitions.

Finally, we apply conflict-based refinement for functional relations. When a newer nugget with the same key $\kappa$ has sufficient evidence ($\geq 2$ sources), we set the current nugget's $t_e$ to the newer nugget's $t_s$, treating this as temporal succession. This heuristic prioritizes recency for single-valued attributes (e.g., CEO) but does not apply to multi-valued relations (e.g., board members).

\begin{algorithm}[t]
\caption{Validity Interval Inference}
\label{alg:infer-validity}
\footnotesize
\Require{nugget $n$ (text $\ell_n$), doc $d$ (time $t_d$), revision history $H$}
\Ensure{$\sigma_n=[t_s,t_e)$}
$T \gets$ \textsc{ExtractTemporalExpressions}$(\ell_n)$\;
$t_s \gets
\begin{cases}
date & \exists(\_,date,\texttt{START})\in T\\
date & \exists(\_,date,\texttt{POINT})\in T\\
t_d  & \text{otherwise}
\end{cases}$\;
$t_e \gets
\begin{cases}
date & \exists(\_,date,\texttt{END})\in T\\
+\infty & \text{otherwise}
\end{cases}$\;
\If{$d$ has history in $H$}{ \tcp*[r]{tighten via revisions}
  $R \gets$ \textsc{GetRevisions}$(d,H)$ \tcp*[r]{time-ordered}
  \For{$i\gets1$ \KwTo $|R|-1$}{
    $r_i,r_{i+1}\gets R[i],R[i+1]$\;
    \If{$\ell_n\in r_i \land \ell_n\notin r_{i+1}$}{
      $t_e \gets \textsc{Earlier}\!\bigl(t_e,\textsc{Time}(r_{i+1})\bigr)$\;
    }
    \If{$\ell_n\notin r_i \land \ell_n\in r_{i+1}$}{
      $t_s \gets \textsc{Later}\!\bigl(t_s,\textsc{Time}(r_{i+1})\bigr)$\;
    }
  }
}
$N_{\neq} \gets \{\,n' \mid \kappa_{n'}=\kappa_n \land \text{value}(n')\neq \text{value}(n)\,\}$\;
\ForEach{$n'\in N_{\neq}$}{ \tcp*[r]{tighten via conflicts}
  \If{$t_s(n')>t_s \land |\textsc{Evidence}(n')|\ge2$}{
    $t_e \gets \textsc{Earlier}\!\bigl(t_e, t_s(n')\bigr)$\;
  }
}
\Return{$[t_s,t_e)$}\;
\end{algorithm}

\begin{algorithm}[t]
\caption{Conflict Detection and State Assignment}
\label{alg:conflict-detection}
\footnotesize
\Require{candidate nugget $n_{\text{new}}$, index $\mathcal{N}$, schema $\mathcal{R}$}
\Ensure{updated index and states $s_\bullet$}
$\kappa \gets$ \textsc{ComputeKey}$(n_{\text{new}})$\;
$N_{\kappa} \gets \{\,n\!\in\!\mathcal{N} \mid \kappa_n=\kappa\,\}$\;
\If{$N_{\kappa}=\emptyset$}{ \tcp*[r]{new key}
  $s_{n_{\text{new}}}\gets\mathsf{Active}$; \Return{$\mathcal{N}\cup\{n_{\text{new}}\}$}\;
}
\ForEach{$n\in N_{\kappa}$}{ \tcp*[r]{exact key match}
  \If{\textsc{JaccardValue}$(n,n_{\text{new}}) \ge 0.85$}{ \tcp*[r]{deduplication}
    $\pi_n \gets \pi_n\cup\pi_{n_{\text{new}}}$; \Return{$\mathcal{N}$} \tcp*[r]{merge evidence}
  }
}
$r \gets$ \textsc{GetRelation}$(\kappa)$\;
$N_{\cap} \gets \{\,n\!\in\!N_{\kappa} \mid \sigma_n\cap\sigma_{n_{\text{new}}}\neq\emptyset\,\}$\;
\If{$N_{\cap}=\emptyset \,\mathbf{or}\, r\in\mathcal{R}_{\text{multi}}$}{ \tcp*[r]{succession / multi-valued}
  $s_{n_{\text{new}}}\gets\mathsf{Active}$; \Return{$\mathcal{N}\cup\{n_{\text{new}}\}$}\;
}
\ForEach{$n\in N_{\cap}$}{ \tcp*[r]{functional + overlap}
  \If{$|\pi_{n_{\text{new}}}|\ge2 \land t_s(n_{\text{new}})>t_s(n)$}{
    $s_n\gets\mathsf{Deprecated}$; $t_e(n)\gets t_s(n_{\text{new}})$; $s_{n_{\text{new}}}\gets\mathsf{Active}$\;
  }
  \ElseIf{$|\pi_n|\ge2 \land t_s(n)>t_s(n_{\text{new}})$}{
    $s_{n_{\text{new}}}\gets\mathsf{Deprecated}$; $t_e(n_{\text{new}})\gets t_s(n)$\;
  }
  \Else{
    $s_n,s_{n_{\text{new}}}\gets\mathsf{Contested}$\;
  }
}
\Return{$\mathcal{N}\cup\{n_{\text{new}}\}$}\;
\end{algorithm}

\subsubsection{Stage 4: Conflict Handling and State Assignment}
Conflict detection is algorithmic, not generative. The LLM extracts facts from documents without cross-document awareness. The system identifies conflicts downstream when multiple nuggets share the same key $\kappa$ but have different values and overlapping validity intervals.

This design provides robustness to LLM inconsistency. A hallucinated fact from a single document lacks the evidence redundancy required to deprecate existing knowledge and is isolated as \textsc{\small Deprecated}. The existing consensus remains \textsc{\small Active} until the new claim accumulates sufficient corroboration.

Following validity inference, we integrate candidates into the index (Algorithm~\ref{alg:conflict-detection}) via deduplication and state assignment:

\paragraph{Deduplication (Merge).} If the candidate matches an existing key $\kappa$, we compare object values. Exact matches or near-duplicates (Jaccard $\geq 0.85$ over character n-grams) are merged. For example, ``Google's CEO Sundar Pichai'' and ``Pichai, leader of Google'' resolve to the same key and similar values, allowing storage as a single nugget with accumulated evidence.

\paragraph{Conflict Detection (Update).} If keys match but values differ (Jaccard $< 0.85$) and the relation is functional, we examine temporal overlap. Non-overlapping intervals indicate succession (e.g., a CEO change), yielding a new \textsc{\small Active} nugget. Overlapping intervals trigger conflict resolution. Multi-valued relations (e.g., board members) accept concurrent values. For functional relations: if the newer candidate has $\geq 2$ independent sources, the existing nugget is marked \textsc{\small Deprecated}; if the existing nugget has superior evidence, the candidate is deprecated; ambiguous cases mark both \textsc{\small Contested}, requiring $\geq 3$ sources for resolution. The cardinality is a per-predicate property of the schema. Legitimate concurrency on a normally-functional role (e.g., a co-CEO arrangement during a leadership transition, or co-chairs on a board) is supported by overriding the predicate's cardinality to multi-valued in a user-supplied schema, rather than being silently misclassified as a conflict. The schema-discovery module described in Section~\ref{sec:method} surfaces such overrides automatically when the input corpus contains multiple concurrent objects on a predicate that the default schema marks functional.

These thresholds were selected empirically, balancing noise reduction against update recall. The thresholds assume source independence; correlated errors (consistent LLM hallucinations or circular reporting) could bypass this safeguard. To mitigate this, changes to frequently accessed nuggets are flagged for human review. \systemname{} includes an annotation tool for manual verification of extraction, normalization, validity, and conflict decisions\footnote{\href{https://github.com/searchsim-org/sigir26-nuggetindex}{https://github.com/searchsim-org/sigir26-nuggetindex}}.

\subsection{Indexing and Retrieval}

The \systemname{} comprises four decoupled components that can be enabled or disabled based on deployment constraints:

\begin{enumerate}[label=(\arabic*),labelindent=0pt,leftmargin=2em,labelsep=0.4em] 
    \item \textbf{Document Store:} A key-value store mapping nugget IDs to full records (text, validity, state, provenance).
    \item \textbf{Metadata Index:} A B-tree~\cite{bayer1970organization} over validity intervals and states. This component is mandatory for governance but requires negligible storage.
    \item \textbf{Sparse Index:} An inverted index storing the canonical nugget text $\ell_n$ for BM25 retrieval~\cite{robertson2009probabilistic}. It requires no heavy inference hardware.
    \item \textbf{Dense Index:} An HNSW~\cite{malkov2018efficient} approximate nearest neighbor structure \text{\small($M=32$, \texttt{efConst}=200)} over nugget embeddings $\mathbf{z}_n$.
\end{enumerate}

This modularity supports flexible deployment. \textit{Hybrid} mode uses all components for high-resource environments. \textit{Lexical-Only} mode omits the dense index for resource-constrained or edge environments (e.g., browser-based search), removing embedding inference while retaining temporal governance.

\subsubsection{Retrieval Algorithm.}
Retrieval for a query $x$ at time $t$ involves a four-step procedure. First, \textbf{view filtering} restricts the search to nuggets in the target view (\textit{active} or \textit{active-plus-contested}) that are valid at time $t$. The metadata index eliminates nuggets where $t \notin \sigma_n$ or where $s_n$ is \textsc{\small Deprecated}. Second, \textbf{candidate retrieval} computes BM25 scores and retrieves approximate nearest neighbors for the query embedding within this filtered set.

Third, \textbf{score fusion} combines these components via $s(x, n) = \alpha \cdot s_{\text{BM25}}(x, n) + \beta \cdot s_{\text{dense}}(x, n)$, with hyperparameters $\alpha$ and $\beta$  determined via grid search on development data. Finally, the system ranks nuggets by the combined score to return the top-$K$ results.

\subsubsection{Handling Contested Nuggets}
Using the \textit{active-plus-contested} view, \systemname{} retrieves contested nuggets and explicitly flags them in the generator prompt. Unlike standard RAG, which may arbitrarily select one side of a contradiction or hallucinate a resolution, the system preserves the dispute structurally. The generator is instructed to encode this uncertainty using the following schema:

\begin{prompt}
\small\ttfamily
Established facts:\\
- [Active nugget 1]\\
- [Active nugget 2]\\
\\
Disputed (sources disagree):\\
- [Contested nugget]: Source A says X, Source B says Y
\end{prompt}

This structured injection lets the generator produce nuanced answers (e.g., ``while generally considered X, some sources claim Y...''). Consequently, the frequency of confident hallucinations decreases.

\section{Experimental Setup}
\label{sec:setup}

We evaluate \systemname{} across five research questions, measuring retrieval effectiveness (RQ1--RQ2), governance accuracy (RQ3--RQ4), and practical deployment (RQ5). This section describes the evaluation framework. Section~\ref{sec:datasets} introduces the datasets, Section~\ref{sec:baselines} the baselines, and Section~\ref{sec:implementation} the implementation details.

\subsection{Research Questions}

\noindent Five research questions guide the evaluation:
\begin{rqlist}
  \item \textbf{RQ1 (Coverage).} Assesses whether \systemname{} improves nugget recall and answer completeness against passage baselines.
  \item \textbf{RQ2 (Temporal Correctness).} Evaluates temporal accuracy and version conflict reduction under corpus evolution.
  \item \textbf{RQ3 (Efficiency).} Determines whether nugget retrieval reduces prompt length without degrading generation quality.
  \item \textbf{RQ4 (Ablations).} Measures the individual contributions of granularity, validity filtering, and lifecycle states.
  \item \textbf{RQ5 (Construction Quality).} Analyzes the accuracy of the automated nugget construction pipeline.
\end{rqlist}

\subsection{Datasets}
\label{sec:datasets}

We use four benchmarks covering static retrieval, temporal reasoning, and multi-hop question answering. To address the lack of nugget-level annotations in standard datasets, we combine existing nuggetized resources with adapted benchmarks.

\subsubsection*{RAVine: Nuggetized MS MARCO (RQ1)}
For static coverage evaluation, we use RAVine~\cite{xu2025ravine}, based on MS MARCO v2.1~\cite{bajaj2016ms,trecRAGcorpus2024msmarcoV21}. The dataset contains 84 queries with 3,182 gold nuggets classified as \textit{vital} (1,602) or \textit{okay} (1,580), linked to 2,500 documents within a corpus of approximately 10 million passages.

\subsubsection*{TimeQA (RQ2, RQ4, RQ5)}
For temporal correctness, we use TimeQA~\cite{chen2021dataset} with 12,183 questions in easy and hard splits. Derived from Wikipedia revision histories, the dataset targets entities with time-varying attributes. The hard split tests temporal disambiguation where answers vary across periods.

\subsubsection*{SituatedQA (RQ2)}
We use the temporal subset of SituatedQA~\cite{zhang2021situatedqa} (12,227 questions). Unlike TimeQA, this dataset evaluates implicit temporal grounding without explicit time references (e.g., ``current president''), testing robustness to implicit signals.

\subsubsection*{MuSiQue (RQ3)}
For multi-hop efficiency, we use MuSiQue~\cite{trivedi2022musique} (47,251 questions with decomposed reasoning chains), measuring token use and latency when nugget retrieval reduces context length.

\subsubsection*{Statistical Considerations}
Although RAVine contains only 84 queries, its 3,182 nuggets ensure statistical significance~\cite{dror2018hitchhiker}. Multiple runs yield narrow 95\% confidence intervals ($\pm$.017 to $\pm$.034) with $p < 0.001$. The combined TimeQA and SituatedQA datasets provide over 24,000 questions for temporal evaluation.

\subsubsection*{Limitations}
A primary limitation is the scarcity of nuggetized benchmarks. Standard resources like Natural Questions~\cite{kwiatkowski2019natural} and HotpotQA~\cite{yang2018hotpotqa} lack nugget-level decomposition. Creating such annotations requires significant manual or verified automated effort. Consequently, RAVine serves as the most rigorous foundation for our coverage claims due to its expert-curated annotations.

\subsection{Baselines}
\label{sec:baselines}

We compare our approach against four baseline categories to separate the effects of granularity, time awareness, and governance.

\subsubsection*{Passage Retrieval}
Three standard retrievers: \emph{BM25-Passage} (sparse lexical matching)~\cite{robertson2009probabilistic}; \emph{Dense-Passage} (DPR-style)~\cite{karpukhin2020dense}; and \emph{Hybrid-Passage} (linear combination of BM25 and dense scores)~\cite{ma2021replication}.

\subsubsection*{Time-Aware Passage Retrieval}
To isolate governance from simple time metadata, we augment the hybrid baseline with temporal logic. \emph{TimeFilter} restricts retrieval to a query time window~\cite{li-croft-2003-time}. \emph{RecencyRerank} applies score decay $s' = s \cdot \exp(-\lambda(t - t_{\text{doc}}))$~\cite{li-croft-2003-time,campos2016gte}. \emph{LatestSnapshot} indexes only the most recent document version~\cite{huwiler2025versionrag}.

\subsubsection*{Proposition Retrieval}
To test granularity without governance, \emph{Proposition-RAG} extracts propositions using the same method as \systemname{} (Stage 1) but indexes them as plain text without validity intervals or lifecycle states~\cite{chen2024dense}. \emph{Proposition-RAG + TimeFilter} adds document-level time filtering~\cite{li-croft-2003-time}.

\subsubsection*{Graph-Based Retrieval}
\emph{GraphRAG}~\cite{knollmeyer2025document} builds an entity graph from documents and retrieves via graph traversal and reranking using the authors' default parameters.

% ------------------------------------------------------------------------------------------------------------------------
% TABLE RESULTS FOR SECTION 05 -- I placed it here to force show it in this page, didn't manage otherwise
% ------------------------------------------------------------------------------------------------------------------------
\begin{figure*}[t]
\centering

% ---------- LEFT: TABLE (60%) ----------
\begin{minipage}[t]{0.60\textwidth}
\caption{Results on nuggetized MS~MARCO ---RAVine~\cite{xu2025ravine}.}
\label{tab:msmarco-results}
\centering

\small
\setlength{\tabcolsep}{2.2pt}
\renewcommand{\arraystretch}{1.08}

\begin{tabular*}{\linewidth}{@{\extracolsep{\fill}}lcccccc@{}}
\toprule
\multirow{2}{*}{System} & \multicolumn{4}{c}{Retrieval} & \multicolumn{2}{c}{Nugget Coverage} \\
\cmidrule(lr){2-5}\cmidrule(lr){6-7}
& nDCG@10 & R@10 & R@20 & R@50 & Nugget R & Vital R \\
\midrule
\multicolumn{7}{@{}l}{\textit{Passage retrieval}} \\
BM25-Passage   & .325 {\footnotesize$\pm$.031} & .091 {\footnotesize$\pm$.013} & .118 {\footnotesize$\pm$.014} & .184 {\footnotesize$\pm$.018} & .166 {\footnotesize$\pm$.017} & .185 {\footnotesize$\pm$.019} \\
Dense-Passage  & .349 {\footnotesize$\pm$.034} & .071 {\footnotesize$\pm$.009} & .098 {\footnotesize$\pm$.012} & .119 {\footnotesize$\pm$.013} & .136 {\footnotesize$\pm$.015} & .149 {\footnotesize$\pm$.017} \\
Hybrid-Passage & .324 {\footnotesize$\pm$.031} & .091 {\footnotesize$\pm$.013} & .118 {\footnotesize$\pm$.014} & .184 {\footnotesize$\pm$.018} & .166 {\footnotesize$\pm$.017} & .185 {\footnotesize$\pm$.019} \\
\midrule
\multicolumn{7}{@{}l}{\textit{Proposition retrieval \footnotesize (atomic, no governance)}} \\
Proposition-RAG & .218 {\footnotesize$\pm$.030} & .062 {\footnotesize$\pm$.012} & .100 {\footnotesize$\pm$.014} & .100 {\footnotesize$\pm$.014} & .144 {\footnotesize$\pm$.018} & .158 {\footnotesize$\pm$.019} \\
\midrule
\multicolumn{7}{@{}l}{\textit{Graph-based retrieval}} \\
GraphRAG & .205 {\footnotesize$\pm$.028} & .064 {\footnotesize$\pm$.010} & .097 {\footnotesize$\pm$.014} & .098 {\footnotesize$\pm$.014} & .142 {\footnotesize$\pm$.017} & .166 {\footnotesize$\pm$.019} \\
\midrule
\multicolumn{7}{@{}l}{\textit{\textsc{NuggetIndex}}} \\
\textsc{NuggetIndex}-Active &
\textbf{.637}$^\dagger${\scriptsize$\pm$.034} &
\textbf{.312}$^\dagger${\scriptsize$\pm$.027} &
\textbf{.353}$^\dagger${\scriptsize$\pm$.028} &
\textbf{.372}$^\dagger${\scriptsize$\pm$.025} &
\textbf{.235}$^\dagger${\scriptsize$\pm$.021} &
\textbf{.272}$^\dagger${\scriptsize$\pm$.024} \\
\bottomrule
\end{tabular*}

\captionsetup{type=table}
\floatnote{Retrieval metrics evaluate the identification of source documents containing gold nuggets. Values represent means across 5 runs (420 queries); $\pm$ denotes 95\% confidence intervals. \textbf{Bold} indicates the best result, while $\dagger$ marks significant improvement over the top passage baseline ($p<0.001$).}
\end{minipage}
\hfill
% ---------- RIGHT: FIGURE (40%) ----------
\begin{minipage}[t]{0.38\textwidth}
% \vspace{0pt}
\caption{Granularity and prompt efficiency.}
\label{fig:granularity}
\centering

\includegraphics[width=.99\linewidth]{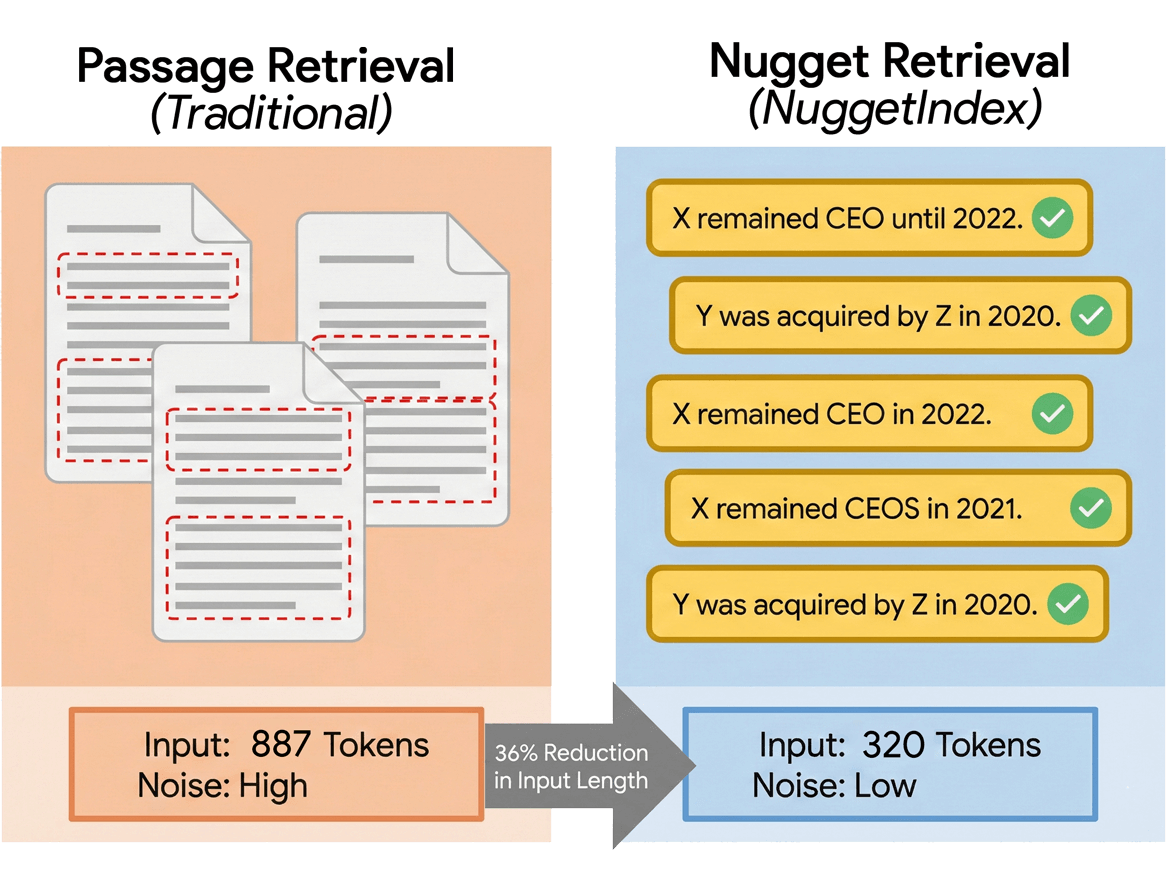}

\captionsetup{type=figure}

\floatnote{Passage retrieval returns long, noisy text blocks; nugget retrieval selects compact facts, reducing prompt noise and input length.}
\end{minipage}

\end{figure*}
% ------------------------------------------------------------

\subsection{Implementation Details}
\label{sec:implementation}

The dense encoder uses Cohere \texttt{\small embed-english-v3}~\cite{cohere-embed-english-v3} (1024 dimensions) on the MS~MARCO v2.1 corpus~\cite{trecRAGcorpus2024msmarcoV21,msmarco}; embedding-model substitutability for retrieval is studied in~\cite{caspari2024beyond}. We retrieve $K{=}20$ nuggets using score-level fusion with weights $\alpha{=}0.4$ (lexical), $\beta{=}0.5$ (dense), and $\gamma{=}0.1$ (metadata). Scores are min--max normalized to $[0,1]$ and combined as $s(d)=\alpha\,s_{\text{lex}}(d)+\beta\,s_{\text{dense}}(d)+\gamma\,s_{\text{meta}}(d)$. The time filter window is $\pm 180$ days with recency decay $\lambda{=}0.001$.

We use \texttt{gpt-4o-mini}~\cite{openai-gpt-4o-mini} as the generator to reflect latency-sensitive production environments. This choice isolates performance gains attributable to governed retrieval rather than the model's parametric knowledge.

Results are means with 95\% confidence intervals from five runs with distinct random seeds. Statistical significance is determined using paired bootstrap tests~\cite{tibshirani1993introduction} (10,000 iterations) with Bonferroni correction~\cite{bonferroni1936teoria} for multiple comparisons. The nugget store is backed by SQLite for the experiments reported here; the storage layer is a thin abstraction and a drop-in PostgreSQL/pgvector backend is supported for production deployments where multi-writer durability and external authentication are required.

\section{Results}
\label{sec:results}

We structure the analysis by research question and keep generator settings fixed within each dataset for comparability.

\subsection{RQ1: Nugget Coverage on Static QA}
\label{sec:results-coverage}

\subsubsection{Motivation.}
Passage-based retrieval returns documents with mixed relevant and irrelevant content, forcing generators to extract facts from noisy context. We hypothesize that nugget-level retrieval improves ranking quality and coverage by directly indexing atomic facts and reducing redundancy through normalization.

\subsubsection{Setting.}
We evaluate on the nuggetized RAVine benchmark~\cite{xu2025ravine}, a static environment with over 10 million passages~\cite{trecRAGcorpus2024msmarcoV21}. We compare \systemname{}-Active against passage baselines (\textit{BM25-Passage}, \textit{Dense-Passage}, \textit{Hybrid-Passage}), \textit{Proposition-RAG}, and \textit{GraphRAG}. \systemname{}-Active retrieves only nuggets with valid intervals at query time and non-deprecated lifecycle state.

Passage baselines map to nuggets via evidence span containment (validated on 500 queries). Metrics include nDCG@10, R@10/20/50, Nugget Recall, and Vital Nugget Recall. Low absolute recall across all systems reflects task difficulty: retrieving documents supporting fine-grained nuggets from 10 million passages.

\subsubsection{Results.}
Table~\ref{tab:msmarco-results} presents retrieval performance. Among passage baselines, \textit{BM25-Passage} and \textit{Hybrid-Passage} achieve nugget recall of .166. \textit{Dense-Passage} yields higher nDCG@10 (.349) but lower nugget recall (.136), indicating dense representations favor semantic similarity over lexical precision for factual matching.

Proposition-RAG and GraphRAG show comparable nugget recall (.144, .142) but lower nDCG@10 (.218, .205). Heuristic decomposition introduces noise, offsetting atomic granularity benefits.

\systemname{}-Active outperforms all baselines. Compared to \textit{BM25-Passage}, it improves nDCG@10 by 96\% (.637), R@10 by 243\% (.312), nugget recall by 42\% (.235), and vital nugget recall by 47\% (.272). All improvements are significant ($p < 0.001$)~\cite{smucker2007comparison}.

Ranking quality improves more than recall. \systemname{} retrieves nuggets directly and maps them to documents, ensuring high relevance in top results. Normalization consolidates duplicates, and governance filters deprecated facts. \systemname{} retrieves approximately one-third of relevant documents (R@20 .353, R@50 .372), improving over the baseline ceiling of 12\%.

% ------------------------------------------------------------
\begin{figure*}[t]
\centering

% ---------- LEFT: TABLE (60%) ----------
\begin{minipage}[t]{0.60\textwidth}
\caption{Results on TimeQA~\cite{chen2021dataset}.}
\label{tab:temporal-results}
% \vspace{0pt}
\centering

\small
\setlength{\tabcolsep}{2.6pt}
\renewcommand{\arraystretch}{1.10}

\begin{tabular*}{\linewidth}{@{\extracolsep{\fill}}lccccc@{}}
\toprule
System
& \thdr{Nugget}{R@20}
& \thdr{Nugget R}{(gen)}
& \thdr{Nugget F1}{(gen)}
& \thdr{Temporal}{Correctness}
& \thdr{Conflict}{Rate} \\
\midrule

\multicolumn{6}{@{}l}{\textit{Passage retrieval}} \\
Hybrid-Passage
& \textbf{.497}{\footnotesize$\pm$.019} & .143{\footnotesize$\pm$.007} & .165{\footnotesize$\pm$.007}
& .840{\footnotesize$\pm$.006} & .161{\footnotesize$\pm$.008} \\
\midrule

\multicolumn{6}{@{}l}{\textit{Time-aware passage retrieval}} \\
Hybrid + TimeFilter
& .002{\footnotesize$\pm$.002} & .111{\footnotesize$\pm$.006} & .121{\footnotesize$\pm$.006}
& .921{\footnotesize$\pm$.005} & .148{\footnotesize$\pm$.008} \\
Hybrid + RecencyRerank
& .002{\footnotesize$\pm$.002} & .110{\footnotesize$\pm$.006} & .121{\footnotesize$\pm$.007}
& .921{\footnotesize$\pm$.005} & .148{\footnotesize$\pm$.008} \\
Hybrid + LatestSnapshot
& .000{\footnotesize$\pm$.000} & .059{\footnotesize$\pm$.009} & .059{\footnotesize$\pm$.009}
& .925{\footnotesize$\pm$.009} & \textbf{.001}{\footnotesize$\pm$.001} \\
\midrule

\multicolumn{6}{@{}l}{\textit{Proposition retrieval \footnotesize(atomic, no governance)}} \\
Proposition-RAG
& .492{\footnotesize$\pm$.019} & .128{\footnotesize$\pm$.008} & .140{\footnotesize$\pm$.008}
& .838{\footnotesize$\pm$.006} & .102{\footnotesize$\pm$.009} \\
Proposition + TimeFilter
& .002{\footnotesize$\pm$.002} & .101{\footnotesize$\pm$.007} & .106{\footnotesize$\pm$.007}
& .930{\footnotesize$\pm$.004} & .114{\footnotesize$\pm$.011} \\
\midrule

\multicolumn{6}{@{}l}{\textit{Graph-based retrieval}} \\
GraphRAG
& .374{\footnotesize$\pm$.018} & \textbf{.271}{\footnotesize$\pm$.015} & \textbf{.277}{\footnotesize$\pm$.015}
& .846{\footnotesize$\pm$.009} & .045{\footnotesize$\pm$.007} \\
\midrule

\multicolumn{6}{@{}l}{\textit{\textsc{NuggetIndex}}} \\
\textsc{NuggetIndex}-Active
& .343{\footnotesize$\pm$.018} & .115{\footnotesize$\pm$.007} & .125{\footnotesize$\pm$.007}
& .931{\footnotesize$\pm$.005} & .072{\footnotesize$\pm$.008} \\
\textsc{NuggetIndex}-Full
& .359{\footnotesize$\pm$.019} & .115{\footnotesize$\pm$.007} & .125{\footnotesize$\pm$.008}
& \textbf{.934}{\footnotesize$\pm$.004} & .112{\footnotesize$\pm$.010} \\
\bottomrule
\end{tabular*}

\captionsetup{type=table}

% \vspace{-4mm}
\floatnote{Temporal Correctness (TC) is the fraction of covered nuggets valid at query time. Conflict Rate is the fraction of answers containing temporally inconsistent facts (lower is better). \textbf{Bold} indicates the best result.}
\end{minipage}
\hfill
% ---------- RIGHT: FIGURE (40%) ----------
\begin{minipage}[t]{0.38\textwidth}
\caption{Contribution to Governance Score.}
\label{fig:ablation}
\centering

\includegraphics[width=\linewidth]{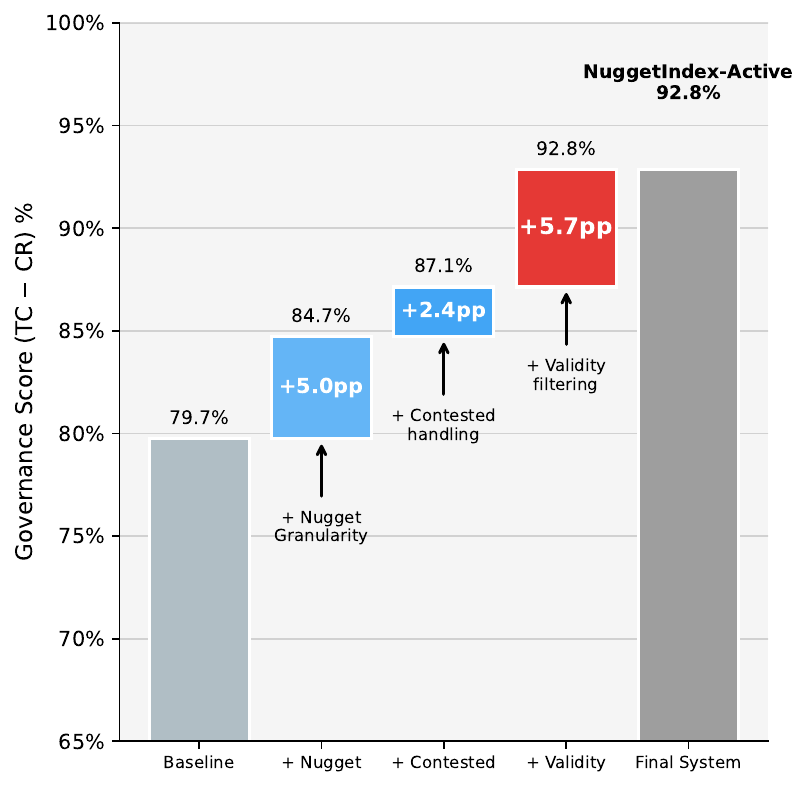}

\captionsetup{type=figure}

\floatnote{Governance Score $=$ Temporal Correctness - Conflict Rate; higher scores indicate greater temporal accuracy and fewer contradictions.}
\end{minipage}

\end{figure*}
% ------------------------------------------------------------

\subsection{RQ2: Temporal Correctness Under Corpus Evolution}
\label{sec:results-temporal}

\subsubsection{Motivation.}
Knowledge bases evolve over time, with facts becoming outdated or contradicted. Standard retrieval systems lack mechanisms to distinguish valid from deprecated content. We hypothesize that governed metadata---validity intervals and lifecycle states---maintains temporal correctness without sacrificing recall, overcoming the trade-off in simple time filtering.

\subsubsection{Setting.}
We evaluate on TimeQA~\cite{chen2021dataset}, targeting entities with time-varying attributes from Wikipedia revision histories. We compare two \systemname{} configurations: \textit{Active} (retrieves only \textit{active} nuggets valid at query time) and \textit{Full} (i.e., \textit{active-plus-contested}, includes contested nuggets with uncertainty prompting). Baselines include \textit{Hybrid-Passage} and three time-aware variants: \textit{TimeFilter} ($\pm$180 days), \textit{RecencyRerank} ($\lambda{=}0.001$), and \textit{LatestSnapshot}. We also include Proposition-RAG (with/without filtering) and GraphRAG.

Metrics include Temporal Correctness (TC, fraction of nuggets valid at query time), Conflict Rate (fraction of answers with inconsistent facts), Nugget R@20, and generation-level Recall/F1 via autograder~\cite{pradeep-etal-2024-autonuggetizer} (89\% human agreement).

\subsubsection{Results.}
Table~\ref{tab:temporal-results} reveals a core trade-off: time-aware filtering improves correctness but collapses recall. \textit{Hybrid-Passage} achieves .840 TC with .497 recall; \textit{TimeFilter} improves TC to .921 but drops recall to .002. \textit{LatestSnapshot} yields lowest conflict rate (.001) but zero recall. Proposition-RAG reduces conflict rates by 37\% (.161 to .102) via atomic decomposition, but TC remains at .838---granularity alone is insufficient. GraphRAG achieves low conflicts (.045) but lower recall (.374).

\systemname{}-Active resolves this trade-off, reaching .931 TC at .343 recall (+9.1pp TC over \textit{Hybrid-Passage} with no recall loss) and cutting conflicts by 55\% to .072. \systemname{}-Full achieves .934 TC with slightly higher conflicts (.112), providing historical context. Handling ``\textit{contested}'' states is key. By flagging conflicting updates as disputed rather than forcing binary decisions, the system allows generators to qualify answers (e.g., ``The exact date is disputed...''), turning potential errors into correct uncertainty statements.

\paragraph{SituatedQA Evaluation.}
On SituatedQA~\cite{zhang2021situatedqa}, \systemname{}-Active achieves .010 conflict rate (96\% reduction vs.\ \textit{Hybrid-Passage} at .230) while preserving .700 recall, confirming governance generalizes beyond revision-based corpora.\footnote{SituatedQA: \href{https://github.com/searchsim-org/sigir26-nuggetindex/tree/main/evaluation/results/rq2/situatedqa}{https://github.com/searchsim-org/sigir26-nuggetindex/situatedqa}}

\subsection{RQ3: Efficiency and Token Economy}
\label{sec:results-efficiency}

\subsubsection{Motivation.}
Large language models costs scale with context length, making prompt efficiency critical for production. Passage retrieval includes redundant content, inflating token counts. We hypothesize that nugget-level retrieval reduces input length while maintaining generation quality, enabling deployment in latency-sensitive or resource-constrained environments.

\subsubsection{Setting.}
We evaluate on TimeQA~\cite{chen2021dataset} (1,000 queries) and MuSiQue~\cite{trivedi2022musique} (2,154 queries). Metrics include build time, build cost (GPT-4o-mini pricing~\cite{openai_pricing}), index size, retrieval latency (P50/P95), median input length, and end-to-end P50 latency.

\subsubsection{Results.}
Table~\ref{tab:efficiency-results} and Figure~\ref{fig:granularity} shows efficiency results. Passage systems require median 887 tokens with negligible build cost. Proposition-RAG reduces to 346 tokens (\$1.40 build cost); GraphRAG achieves 54 tokens through aggressive compression that may lose context. \systemname{} requires 320 tokens---64\% reduction over passage retrieval---with higher information density than propositions.

\systemname{} matches Passage retrieval latency (0.8ms P50 Active; 1.2ms Full vs. 1–2ms). End-to-end latency (1{,}269ms P50) equals \textit{BM25-Passage}, as reduced generation offsets retrieval overhead. The index for 56,990 nuggets requires only 9.3MB, enabling resource-constrained deployment with sub-millisecond latency.

\paragraph{MuSiQue Evaluation.}
On MuSiQue~\cite{trivedi2022musique}, \systemname{}-Active reduces tokens by 82\% (339 vs.\ 1{,}884), confirming that nugget-level retrieval maintains its advantage on multi-hop reasoning.\footnote{MuSiQue: \href{https://github.com/searchsim-org/sigir26-nuggetindex/tree/main/evaluation/results/rq3/musique}{https://github.com/searchsim-org/sigir26-nuggetindex/musique}}

\begin{table}[t]
\caption{Efficiency comparison on TimeQA (1,000 queries).}
\label{tab:efficiency-results}
\small
\centering

\setlength{\tabcolsep}{2.2pt}
\renewcommand{\arraystretch}{1.08}

\noindent\makebox[\columnwidth][c]{%
\begin{tabular*}{\columnwidth}{@{}@{\extracolsep{\fill}}lcccccc@{}}
\toprule
System & \thdr{Build}{Time (h)} & \thdr{Build}{Cost (\$)} & \thdr{Index}{Size (GB)} &
\thdr{Retrieval}{P50/P95} & \thdr{Input}{Length} & \thdr{E2E}{P50} \\
\midrule
BM25-Passage          & .001 & 0    & .016 & 2/5 & 887 & 1{,}269 \\
Hybrid-Passage        & .002 & .04 & .014 & 1/3 & 887 & 1{,}283 \\
\midrule
Proposition-RAG              & .004 & 1.40 & .014 & 1/4 & 346 & 1{,}274 \\
\midrule
GraphRAG              & .007 & 1.30 & .003 & 0/1 & \textbf{54}  & \textbf{1{,}032} \\

\midrule
\small\systemname{}-Active  & .002 & 1.32 & .009 & 1/3 & \uline{320} & \uline{1{,}269} \\
\small\systemname{}-Full    & .003 & 1.32 & .009 & 1/4 & 320 & 1{,}305 \\
\bottomrule
\end{tabular*}%
}

\end{table}

\subsection{RQ4: Ablation Studies}
\label{sec:results-ablations}

\subsubsection{Motivation.}
\systemname{} comprises: nugget granularity, validity filtering, lifecycle states, contested handling, and normalization. Understanding individual contributions guides design decisions and identifies which components justify their complexity. We quantify the trade-off between retrieval coverage and temporal integrity introduced by each governance mechanism.

\subsubsection{Setting.}
We analyze 500 queries from TimeQA~\cite{chen2021dataset}, prioritizing passages with \textsc{\small Deprecated} or \textsc{\small Contested} nuggets. We use \systemname{} with BM25 retrieval to isolate governance effects.

We define \textit{Governance Score} as Temporal Correctness minus Conflict Rate (1.0 = perfect). Ablations disable individual components: validity filtering (retrieving all nuggets regardless of interval), lifecycle states (treating all nuggets as \textsc{\small Active}), contested handling (treating disputed nuggets as \textsc{\small Active} without uncertainty markers), normalization, and nugget granularity (replacing with passage-level retrieval). We also remove all governance components.

\begin{table}[t]
\caption{Ablation study on TimeQA~\cite{chen2021dataset} (500 queries).}
\label{tab:ablation-results}
\small
\centering
\setlength{\tabcolsep}{3pt}
\renewcommand{\arraystretch}{1.05}

\begin{tabular*}{\columnwidth}{@{\extracolsep{\fill}}p{0.4\columnwidth}cccc@{}}
\toprule
Configuration &
\makecell{Nugget\\R@20} &
\makecell{$\Delta$R} &
\makecell{Governance\\Score$^\dagger$} &
\makecell{$\Delta$Gov} \\
\midrule
\textsc{NuggetIndex}-Active & .343 & — & \textbf{.928} & — \\
\midrule
\quad $-$ validity filtering   & .461 & +11.8 & .833 & $-$9.5 \\
\quad $-$ lifecycle states     & .343 & —     & .928 & — \\
\quad $-$ contested handling   & .352 & +0.9  & .888 & $-$4.0 \\
\quad $-$ normalization        & .341 & $-$0.2 & .928 & — \\
\quad $-$ nugget granularity   & .378 & +3.5  & .845 & $-$8.3 \\
\quad $-$ all governance       & .482 & +13.9 & .797 & $-$13.1 \\
\bottomrule
\end{tabular*}

\floatnote{\footnotesize $^\dagger$Governance Score = Temporal Correctness $-$ Conflict Rate.}
\end{table}

\begin{table}[t]
\caption{Retrieval mode ablation on TimeQA ($N=200$).}
\label{tab:retrieval-modes}
\small
\centering

\setlength{\tabcolsep}{3pt}
\renewcommand{\arraystretch}{1.05}

\begin{tabular*}{\columnwidth}{@{\extracolsep{\fill}}p{0.4\columnwidth}cccc@{}}
\toprule
\thdr{Mode}{}
& \thdr{R@10}{}
& \thdr{nDCG}{}
& \thdr{Temporal}{Correctness}
& \thdr{Latency}{} \\
\midrule
Hybrid (Default) & \textbf{.305} & \textbf{.102} & 1.00 & 12.3ms \\
Semantic-Only    & .260 & .083 & 1.00         & 11.5ms \\
Lexical-Only     & .282 & .092 & 1.00         & \textbf{0.4ms} \\
\midrule
\multicolumn{5}{@{}p{\columnwidth}@{}}{\textit{\small Ablation: No Temporal Filter (High Recall / Low Correctness)}} \\
Lexical-Only & \textbf{.372} & .109 & .899 & \textbf{0.3ms} \\
Hybrid       & .357          & \textbf{.110} & .880 & 12.3ms \\
\bottomrule
\end{tabular*}

\floatnote{\footnotesize \textbf{Bold} indicates best in class. Lexical-Only removes embedding costs entirely.}
\end{table}

\subsubsection{Results.}
Table~\ref{tab:ablation-results} and Figure~\ref{fig:ablation} present component contributions. Removing \textit{validity filtering} causes the largest shift: recall increases by 11.8 percentage points while Governance Score decreases by 9.5pp (.928 to .833), indicating the system prioritizes factual currency over coverage. \textit{Lifecycle states} show no measurable impact---TimeQA contains few deprecated facts (84 of 69,903), limiting evaluation of this feature. Removing \textit{contested handling} decreases Governance Score by 4.0pp and increases recall by 0.9pp, showing this component effectively filters contradictory information. Disabling \textit{normalization} does not affect Governance Score.

Replacing nuggets with passages increases recall by 3.5pp but drops Governance Score by 8.3pp. Removing all governance yields +13.9pp recall but $-$13.1pp Governance Score (to .797). Overall, the system trades coverage for temporal accuracy.

\paragraph{Retrieval Mode Ablation.}
While the previous ablation examines governance components, we also analyze retrieval architecture by decoupling sparse and dense retrieval (Table~\ref{tab:retrieval-modes}). \textit{Hybrid} achieves 30.5\% recall with perfect temporal correctness. Notably, \textbf{sparse retrieval outperforms dense} (.282 vs.\ .260 recall), suggesting that for atomic nuggets, precise lexical matching of entities and relations is more effective than broad semantic similarity from embeddings. The \textit{Lexical-Only} mode emerges as a highly efficient deployment option, retaining 92\% of hybrid recall while reducing P50 latency by a factor of 30 (12.3ms to 0.4ms), confirming governed retrieval is feasible for edge environments without embedding inference.

\subsection{RQ5: Construction Quality}
\label{sec:results-intrinsic}

\subsubsection{Motivation.}
\systemname{} effectiveness depends on the quality of its automated construction pipeline. Errors in extraction, normalization, validity inference, or conflict detection propagate to retrieval and generation. Evaluating each stage establishes confidence in the system's foundations.

\subsubsection{Setting.}
We evaluate using 500 samples with ground truth and manually annotated data from TimeQA~\cite{chen2021dataset}.

\textbf{Extraction}: 500 nuggets from 124 passages, stratified by property type. Two annotators independently identified atomic facts (blind to system output) then evaluated whether each extracted nugget was correct and atomic. Metrics: (fraction of extracted nuggets that are valid facts), Recall (fraction of identified facts extracted), and Atomicity (fraction of nuggets containing exactly one fact). Inter-annotator agreement: $\kappa$=0.69 (substantial~\cite{landis1977measurement}).

\textbf{Normalization}: 500 entity pairs measuring Alias Resolution Recall (ability to link equivalent mentions) and False Merge Rate (distinct facts incorrectly merged). Inter-annotator: $\kappa$=0.97.

\textbf{Validity inference} (Algorithm~\ref{alg:infer-validity}): TimeQA nuggets with ground-truth temporal validity windows. Metrics: Start Time Accuracy ($\pm$30 days), End Time Detection Recall (ability to identify when facts end), End Time Accuracy (precision when detected).

\textbf{Conflict detection} (Algorithm~\ref{alg:conflict-detection}): state-change decisions on subject-predicate keys ($\kappa$=0.73, substantial). Metrics: Deprecation Precision, Contestation Appropriateness, Missed Conflict Rate.

We release the manually annotated data to support research on nugget extraction and conflict detection evaluation.\footnote{Annotation: \href{https://github.com/searchsim-org/sigir26-nuggetindex/tree/main/annotation_tool/data}{https://github.com/searchsim-org/sigir26-nuggetindex/annotation\_tool}}

\begin{table}[tbp]
\caption{Construction pipeline quality.}
\label{tab:intrinsic-results}
\centering
\small
\setlength{\tabcolsep}{2pt}
\renewcommand{\arraystretch}{0.95}

\begin{tabular*}{\columnwidth}{@{\extracolsep{\fill}}p{0.28\columnwidth}p{0.48\columnwidth}c@{}}
\toprule
Stage & Metric & Value \\
\midrule
\multirow{2}{*}{Extraction}
  & Precision & .860 \\
  & Atomicity & .943 \\
\midrule
\multirow{2}{*}{Normalization}
  & Alias resolution recall & .600 \\
  & False merge rate & .016 \\
\midrule
\multirow{3}{*}{Validity inference}
  & Start time accuracy & .868 \\
  & End time detection recall & .667 \\
  & End time accuracy & .955 \\
\midrule
\multirow{3}{*}{Conflict detection}
  & Deprecation precision & 1.000 \\
  & Contestation appropriateness & 1.000 \\
  & Missed conflict rate & .000 \\
\bottomrule
\end{tabular*}

\floatnote{\footnotesize Inter-annotator agreement: extraction atomicity $\kappa$=.69, normalization $\kappa$=.97. Higher is better for all metrics except false merge rate and missed conflict rate.}
\end{table}

\subsubsection{Results.}
Table~\ref{tab:intrinsic-results} presents results. \textbf{Extraction} achieves .860 precision and .943 atomicity, indicating most nuggets contain a single fact. \textbf{Normalization} yields .600 alias recall with .016 false merge rate, prioritizing precision to prevent incorrect conflict detection.

\textbf{Validity inference} identifies start times well (.868); end times are accurate when found (.955) but the recall is lower (.667) due to facts ending without textual cues---revision-based inference mitigates this vs. text-only methods. \textbf{Conflict detection} achieves perfect deprecation precision 1.000 with no missed conflicts.

We quantified error propagation through an oracle experiment on 500 queries by replacing extracted nuggets with human-verified data. This improved recall by 3.7 percentage points and temporal correctness by 4.5 percentage points, demonstrating that the system remains effective despite extraction imperfections.

\subsection{Multi-Hop Reasoning}
\label{sec:results-multihop}

\subsubsection{Motivation.}
Multi-hop questions require aggregating information across multiple documents, challenging retrieval systems to identify and chain relevant facts. While passage-based retrieval may capture broad context, nugget retrieval surfaces atomic facts as reasoning steps. We evaluate whether \systemname{} supports  multi-hop reasoning and how performance scales with corpus size.

\subsubsection{Setting.}
We evaluate on HotpotQA~\cite{yang2018hotpotqa} in two settings: \textit{closed-corpus} (query-specific contexts, $\sim$1,000 passages) and \textit{pooled-corpus} (500 queries combined, $\sim$5,000 passages) to simulate open-domain retrieval. Metrics: Intermediate Recall (coverage of all supporting passages required for reasoning), Final Recall (coverage of the answer-bearing passage), Success Rate (fraction of queries where the system retrieves sufficient context). We compare Hybrid-Passage, Proposition-RAG, GraphRAG, and \systemname{}-Active.

\begin{table}[t]
\caption{Multi-hop QA results on HotpotQA~\cite{yang2018hotpotqa}.}
\label{tab:multihop-results}
\centering
\small
\setlength{\tabcolsep}{1.8pt}
\renewcommand{\arraystretch}{1.02}

\begin{tabular}{@{}lccc@{\hspace{3pt}}ccc@{}}
\toprule
& \multicolumn{3}{c}{\textbf{Closed-corpus (1K)}} &
  \multicolumn{3}{c}{\textbf{Pooled-corpus (5K)}} \\
\cmidrule(lr){2-4}\cmidrule(lr){5-7}
System & Inter.\ R & Final R & Success & Inter.\ R & Final R & Success \\
\midrule
Hybrid-Passage              & \textbf{.890} & \textbf{.805} & \textbf{1.000} & \textbf{.855} & \textbf{.775} & \textbf{.990} \\
Proposition-RAG             &         .800 &          .730 &           .970 &          .775 &          .690 &          .970 \\
GraphRAG                    &         .690 &          .555 &           .930 &          .645 &          .480 &          .920 \\
\midrule
\textsc{NuggetIndex}-Active  & \underline{.850} &\underline{.775} & \underline{.990} & \underline{.790} & \underline{.715} & \underline{.980} \\
\bottomrule
\end{tabular}

\floatnote{\footnotesize Intermediate recall (Inter. R) measures coverage of all supporting passages. Final recall (Final R) measures coverage of the answer-bearing passage.}
\end{table}

\subsubsection{Results.}
Table~\ref{tab:multihop-results} presents results. In \textit{closed-corpus}, Hybrid-Passage achieves .890 intermediate recall. Proposition-RAG drops to .800 as sentence-level splitting limits context, while GraphRAG underperforms at .690 due to entity-centric extraction missing relational facts. \systemname{}-Active remains competitive at .850, validating atomic fact aggregation effectiveness.

In \textit{pooled-corpus}, recall drops for all systems as the search space grows, but \systemname{}-Active is most robust, reaching .790 intermediate recall vs.\ .775 (Proposition-RAG) and .645 (GraphRAG).

Hybrid-Passage outperforms \systemname{}-Active by 1.5--4.0pp on this static benchmark. HotpotQA passages are human-curated paragraphs that co-locate the multi-hop evidence, so lexical matching alone recovers most of them in one pass. The corpus is also static, with no temporal labels, no rename events, and no functional-predicate conflicts. Validity filtering and contestation cannot help when there is nothing to filter. \systemname{} therefore pays the cost of decomposition without the benefits of governance. We expect the gap to invert on a dynamic variant of HotpotQA in which a fraction of supporting passages is replaced by stale duplicates with revised source dates. The temporal-correctness gains in Section~\ref{sec:results-temporal} (.931 vs .840) and the conflict-detection precision in Section~\ref{sec:results-intrinsic} are the complementary best case for the governance signals.

\subsection{Summary of Findings}

The results yield five main conclusions. First, nugget-level retrieval outperforms Passage baselines, improving nugget recall by 42\% and vital nugget recall by 47\% (RQ1). Second, governed metadata increases temporal correctness by 9.1 percentage points (.931 vs .840) and reduces conflict rates by 55\% compared to strong Passage baselines, avoiding the recall loss typical of time-filtered methods (RQ2). Third, the compact nugget format reduces generator input by 64\% (median 887 to 320 tokens), while the 9.3MB index supports sub-millisecond latency suitable for resource-constrained environments (RQ3). Fourth, validity filtering is the primary driver of temporal accuracy, and sparse retrieval outperforms dense methods on atomic nuggets, enabling edge deployment without embedding inference (RQ4). Fifth, the automated pipeline demonstrates .860 extraction precision and .943 atomicity; oracle experiments indicate that imperfect governance remains effective, as human verification yields only marginal improvements of 3.7--4.5 percentage points (RQ5).

\section{Discussion and Limitations}
\label{sec:discussion}

The performance gains stem from separating fact validity from document timestamps. Standard time-filtering operates at document level, forcing a trade-off between correctness and recall. \systemname{} resolves this by governing individual facts, filtering deprecated content while preserving valid historical information.

Structurally, \systemname{} balances unstructured retrieval with formal knowledge bases. It avoids rigid ontologies like Wikidata while offering greater transparency than graph-based methods. The modular index architecture enables deployment flexibility: hybrid mode for maximum recall, lexical-only for edge environments.

For corpora whose freshest source predates the query time by more than a configurable threshold (default: one year), an optional freshness-fallback layer issues a single web-search query against an external provider (Tavily, Serper, Exa, or Brave)~\cite{zerhoudi2026owlerlite} and ingests the returned snippets through the same extractor and conflict pipeline; the fallback fires only when the existing store has no \textsc{Active} fact valid at the query time, and the retrieved nuggets are tagged with a distinct provenance type so downstream filters can require corroboration before promoting a fallback claim to \textsc{Active}.

\paragraph{Cost and reproducibility.}
The construction pipeline relies on an LLM extractor, and the cost is one-time and amortized. Table~\ref{tab:efficiency-results} reports offline build cost on TimeQA at \$1.32 with \texttt{gpt-4o-mini}~\cite{openai-gpt-4o-mini}, comparable to Proposition-RAG (\$1.40) and GraphRAG (\$1.30). At runtime, \systemname{} retrieves in 0.8\,ms (P50) and reduces the median generator input by 64\% (887 to 320 tokens, RQ3). On a 47K-query workload such as MuSiQue, this saving dwarfs the build cost within hours of serving. The dependency on \texttt{gpt-4o-mini} is at the extractor boundary only. The retrieval and governance layers are pure Python over SQLite, and adopters can swap in any OpenAI-compatible endpoint (open-weights or self-hosted) via the released \texttt{LLMConfig} without changing the pipeline.

\paragraph{Limitations.}
Validity inference and entity normalization remain primary weaknesses. End time detection achieves only 0.667 recall when facts expire without textual cues. Normalization errors cause missed conflicts when distinct mentions resolve to different keys. The Jaccard-based deduplication ($\ge 0.85$) may miss semantic equivalence in complex phrasing, and exact key matching fails when normalization errors occur. We plan to incorporate embedding-based deduplication, deeper knowledge base linking for entity resolution, improved end-time detection via cross-document signals, and streaming updates for real-time corpus maintenance.

\section{Conclusion}
\label{sec:conclusion}

This work addresses the mismatch between nugget-based evaluation and static retrieval units in RAG systems. We introduce \systemname{}, a retrieval index where atomic facts are stored as managed records with validity intervals, lifecycle states, and provenance. Filtering by these properties prevents propagation of outdated or conflicting information.

Experiments across four benchmarks demonstrate 42\% improvement in nugget recall, 9.1pp increase in temporal correctness, and 55\% reduction in conflict rates---without the recall collapse of time-filtered baselines. The compact format reduces generator input by 64\% and yields lightweight indices (9.3MB for 56,990 nuggets) suitable for edge deployment. Oracle experiments confirm that even imperfect automated governance provides most of the benefit, with human verification yielding only marginal gains.

% \begin{acks}

% \noindent
% This work has received funding from the European Union's Horizon Europe research and innovation program under grant agreement No. 101070014 (OpenWebSearch.EU), and the Bavarian State Ministry of Economic Affairs, Regional Development, and Energy (StMWi).

% \end{acks}

% ---------- Bibliography ----------
\bibliographystyle{ACM-Reference-Format}
\bibliography{sample-base}

\end{document}